 \documentclass[preprint,floats,aps,epsfig,nofootinbib,amssymb]{revtex4}

\usepackage{slashed}
\usepackage{slashed}
\usepackage{graphicx,color}
\usepackage{epsfig}
\usepackage{subfigure}
\usepackage{epsfig}
\usepackage{dcolumn}
\usepackage{bm}
\usepackage{color}

\def\lsim{\mathrel{\rlap{\lower4pt\hbox{\hskip1pt$\sim$}}
    \raise1pt\hbox{$<$}}}         
\def\gsim{\mathrel{\rlap{\lower4pt\hbox{\hskip1pt$\sim$}}
    \raise1pt\hbox{$>$}}}         
\def\R{\mathbb {S}}
\def\C{\boldsymbol {\Theta}}

\def\U{\mathbf {S}}

\begin{document}

\vspace*{-5.8ex}
\hspace*{\fill}{ACFI-T15-22}

\vspace*{+3.8ex}

\title{Symmetry Behind the 750 GeV Diphoton Excess}

\author{Wei Chao}
\email{chao@physics.umass.edu}

 \affiliation{ Amherst Center for Fundamental Interactions, Department of Physics, University of Massachusetts-Amherst,
Amherst MA 01003 United States }

\vspace{3cm}

\begin{abstract}

A $750$ GeV resonance has been observed at the Run 2 LHC in the diphoton channel.  
In this paper, we explain this resonance as a CP-even scalar, $\R$, that triggers the spontaneous breaking of local $U(1)_B$ or $U(1)_{B+L}$ gauge symmetries.
$\R$ couples to gluon and photon pairs at the one-loop level, where particles running in the loop are introduced to  cancel anomalies.
And the gluon fusion is the dominate production channel of $\R$ at the LHC. 
The model contains a scalar dark matter candidate, stabilized by the new gauge symmetry. 
Our study shows that both the observed production cross section at the LHC and the best fit  decay width of $\R$ can be explained in this  model without conflicting with any other experimental data. 
Constraints on couplings  associated with $\R$ are studied, which show that $\R$ has a negligible mixing with the standard model Higgs boson but sizable coupling with the dark matter. 
\end{abstract}

\maketitle
\section{Introduction}

Both the ATLAS~\cite{ATLAS13} and CMS~\cite{CMS13} collaborations have observed a resonance  about  $750~{\rm GeV}$ in the diphoton channel.  
It may be just a statistical fluctuation, but absolutely deserves a systematic study of new physics explanations. 
According to the Landau-Yang theorem~\cite{Landau:1948kw,Yang:1950rg}, which uses the general principle of  invariance under rotation and inversion to derive selection rules governing decays of  a particle into diphoton, this new resonance can only be colorless spin-0 or spin-2 bosonic states, whose signal at the LHC is
\begin{eqnarray}
\sigma (pp\to {\R}\to \gamma \gamma ) = {2J+1 \over  s}  { \Gamma_{tot}^{\R} \over M}\left[ C_{gg}^{} {\rm BR } (\R\to gg ) + \sum_q C_{q\bar q}  {\rm BR }(\R\to q\bar q)\right] {\rm BR} (\R\to \gamma \gamma) \; , \label{basic}
\end{eqnarray}
 where $\sqrt{s}$ is the centre-of-mass energy,  $J$ and $M$ are the spin and mass of $\R$ respectively,  $ C_{gg}$ and $C_{q\bar q}$ are dimensionless partonic integrals, ${\rm BR} (\R\to XX)$  and $\Gamma_{\rm tot}^\R$ are the branching ratio and total decay rate of $\R$. 
A first glance of  Eq. (\ref{basic}) shows many possible interactions of the new resonance could explain this anomaly. 
But one needs to explain why the Run-I LHC only saw little excesses~\cite{ATLAS8,CMS8} in the meanwhile. 
From this point of view heavy flavors and/or gluon  fusion production of the resonance are favored, since the luminosity ratio, $r=\sigma_{\rm 13 TeV}/\sigma_{\rm 8 TeV}$,  ranges from $2.5\sim 4.1 $ for light flavors to $4.8\sim 5.4$ for heavy flavors and gluon~\cite{Franceschini:2015kwy}.  
Considering  $C_{gg}/C_{x\bar x}\sim {\cal O } (20\sim 100)$, where $x$ represents  heavy flavors, as well as constraints arising from searches of dijet  at the Run-1 LHC, it turns out the resonance is more likely produced at the LHC via the gluon fusion.

In this paper we assume the resonance is a scalar $\R$.
To explain the $750~{\rm GeV}$ diphoton excess, $\R$ is likely coupled to both photon and gluon pairs. 
The effective interactions of $\R$ with the gluon and photon can be written as
\begin{eqnarray}
\delta { \cal L} \ni\C_{\gamma \gamma }\R {\rm F}_{\mu\nu} {\rm F}^{\mu \nu }+\tilde{\C}_{\gamma \gamma }\R {\rm F}_{\mu\nu} \tilde{{\rm F}}^{\mu \nu }  +  \C_{gg} \R {\rm G}^{a \mu \nu} {\rm G}_{\mu \nu}^a +\tilde{\C}_{gg} \R {\rm G}^{a \mu \nu} \tilde{{\rm G}}_{\mu \nu}^a  \label{effective}
\end{eqnarray}
where $\C_{aa}$ and $\tilde{\C}_{aa}$ are CP-conserving and CP-violating effective couplings respectively, which have inverse energy scale.  
Since CP-violating effective couplings is constrained by the non-observation of the permanent electric dipole moments,  CP-conserving effective couplings are more convenient and less constrained explanation. 
There are already many explanations of the excess~\cite{Angelescu:2015uiz,Backovic:2015fnp,Bellazzini:2015nxw,Buttazzo:2015txu,DiChiara:2015vdm,Ellis:2015oso,Franceschini:2015kwy,Gupta:2015zzs,Harigaya:2015ezk,Higaki:2015jag,Knapen:2015dap,Low:2015qep,Mambrini:2015wyu,McDermott:2015sck,Molinaro:2015cwg,Nakai:2015ptz,Petersson:2015mkr,Pilaftsis:2015ycr,Dutta:2015wqh,Cao:2015pto,Matsuzaki:2015che,Kobakhidze:2015ldh,Martinez:2015kmn,Cox:2015ckc,Becirevic:2015fmu,No:2015bsn,Demidov:2015zqn,Chao:2015ttq,Fichet:2015vvy,Curtin:2015jcv,Bian:2015kjt,Chakrabortty:2015hff,Agrawal:2015dbf,Csaki:2015vek,Falkowski:2015swt,Aloni:2015mxa,Bai:2015nbs,Chao:2015nac,Chao:2016mtn,Chao:2016aer}\footnote{For studies of new physics in the diphoton channel, see, Refs.~\cite{Jaeckel:2012yz,Carmi:2012in}.}, some of which start from Eq.~(\ref{effective}) using the bottom-up strategy while others start with concrete models using the top-down strategy. 

Although many models can explain this excess, one is still not clear with the physics behind this resonance. 
Is this scalar a fundamental particle?  
Does it have anything to do with the symmetry breaking? 
Why this neutral colorless scalar couples to gluon and photon? 
How does it interact with the standard model (SM) Higgs boson?  In this paper we will provide a  concrete explanation to these questions. 
We work in the framework of the SM extended with a local $U(1)_{X}$ gauge symmetry, where $X$ can be local baryon number   $B$ or $B+L$ symmetry, that is spontaneously broken as $\R$ gets nonzero vacuum expectation value (VEV).   
New vector-like quarks are introduced to cancel anomalies, which have Yukawa interactions with $\R$  that generate nonzero fermion masses as the $U(1)_X$ symmetry is broken. 
The same Yukawa interactions induce effecive interactions of $\R$ with gluon-gluon and diphoton at the one-loop level. 
A scalar dark matter, stabilized by the new U(1) symmetry, should be introduced to avoid the problem of long lived charged/colored particles.
We find this model can efficiently explain the $750~{\rm GeV}$  diphoton excess without conflicting with any other experimental data. 
%
%
%
$\R$ can mixies with the SM Higgs boson, but the mixing angle is constrained to be very much small, ${\cal O} (10^{-2})$, which means this resonance is not able to trigger the strongly first order electroweak phase transition required by the electroweak baryogenesis mechanism. 
Commendably, all particles relevant to the explanation of the resonance has their own duty in this model.  
We take it as a prototype of the complete theory behind the 750 GeV diphoton excess.

The remaining of this paper is organized as follows: We briefly describe the  model in section II and study  its constraints  in section III. 
Section IV is devoted to investigate the 750 GeV diphoton excess. 
The last part is concluding remarks.

\begin{table}[htbp]
\centering
\begin{tabular}{ccc||ccr}
\hline SM particles & $G_{SM} $ & $U(1)_X$ & BSM  particles&  $G_{SM}  $ & $U(1)_X$ \\
\hline
$(u,~d)_L$ & $(3,~2, ~1/6)$  & m  &$\psi^{\rm D}_L $  &$(3,~2,~ 7/6) $ & b\\
$u_R$ & $(3,~1, ~2/3)$  & m&$\psi^{\rm D}_R $  &$(3,~2,~ 7/6) $ & -b \\
$d_R$ & $(3,~1, ~-1/3)$  & m &$\psi_{1R}^r $  &$(3,~1, ~5/3) $ & b\\
$(\nu,~e)_L$ & $(1,~2, ~1/2)$  & k &$\psi_{2R}^r $  &$(3,~1, ~2/3) $ & b\\
$e_R$ & $(1,~1, ~-1)$  & k &$\psi_{1L}^l $  &$(3,~1, ~5/3) $ & -b\\
$\nu_R$ & $(1,~1, ~0)$  & k&$\psi_{2L}^l $  &$(3,~1, ~2/3) $ & -b \\
$H$ & $(1,~2,~{1/2})$ & 0 & $\R$ & $(1,~1,~0)$ & -2b \\
\hline 
\end{tabular}
\caption{ Quantum numbers of  fields under  the local gauge symmetries $ G_{SM }\times U(1)_X$, where $G_{SM}=SU(3)_C\times SU(2)_L \times U(1)_Y $. Notice that extra quarks  are not VL with
respect to the $U(1)_X$ because pure VL fermions would not create anomalies.  }\label{aaa}
\end{table}

\section{The model}

In the SM, both the baryon number and the lepton number are accidental global symmetries.  
According to Sakharov~\cite{Sakharov:1967dj}, the baryon number ($B$) must be broken to have a matter-antimatter asymmetric Universe.  
Lepton number ($L$) is  expected to be broken to have Majorana neutrino masses. 
In this paper we take $B$ or $B+L$  as a spontaneously broken gauge symmetry. 
We start with the construction of a general  local $U(1)_X$ gauge symmetry then get back to the gauged $B$ or $B+L$ symmetry as a special case. 
Particle contents and their representations under the $SU(3)_C\times SU(2)_L \times U(1)_Y\times U(1)_X$ are listed in the Table.~\ref{aaa}\footnote{Our particle contents are similar to these Refs~\cite{FileviezPerez:2010gw,FileviezPerez:2011pt,Duerr:2013dza} but charges of fields are different, because we study anomaly cancellations of the $U(1)_{xL+yB}$ gauge symmetry. }.  
Note that new fermions are  VL only with respect to the  SM gauge group, not VL with respect to the $U(1)_X$, because pure VL fermions would not create any anomalies.
The global $SU(2)_L$ anomaly \cite{globalsu2} requires the even number of fermion doublets, which is automatically satisfied.
The absence of axial-vector anomalies \cite{avector1,avector2,avector3} and the gravitational-gauge anomaly \cite{anog1,anog2,anog3} requires that certain sums of the $U(1)^\prime$ charges vanish.  
These anomaly-free conditions finally turns to be
\begin{eqnarray}
3m+k+2x b=0 \; , \label{master}
\end{eqnarray}
where $x$ is the number of families of new fermions.
The eq. (\ref{master}) is the master equation of  anomaly cancellations.  
The weak hypercharge of newly introduced fermions are totally free. 
One has following four interesting scenarios arising from eq. (\ref{master}) by setting $x=1$:
\begin{itemize}
\item  $b=0$, $k=-1$ and $m=1/3$ corresponds to the famous $U(1)_{B-L}$ symmetry. 
\item  $k=0 $, $m=1/3$ and $b=-1/2 $  corresponds to the $U(1)_B$ gauge symmetry.
\item  $m=0$, $k=1$ and $b=-1/2$ corresponds to the $U(1)_L$ gauge symmetry.
\item  $m=1/3$, $k=1$ and $b=-1$ corresponds to the $U(1)_{B+L}$ gauge symmetry.  
\end{itemize}
$U(1)_{B-L}$ gauge symmetry was well-studied, while the phenomenology of $U(1)_{B+L}$ gauge symmetry is not studied before.
Anomaly cancellations of  a $U(1)_{xB+yL}$ gauge symmetry was first proposed in Ref.~\cite{Chao:2010mp}.
$U(1)_L \times U(1)_B$  gauge symmetries were studied in Refs.~\cite{FileviezPerez:2010gw,FileviezPerez:2011pt,Duerr:2013dza}. 
In the following we take the $U(1)_X$ as $U(1)_B$ or $U(1)_{B+L}$ gauge symmetry, and study its possible explanation of the 750 GeV diphoton excess. 
New fermions have following Yukawa interactions:
\begin{eqnarray}
{\cal L}_{\rm Y} &\supset&  \sum_{i=1}^2  y_\psi^{i} \overline{\psi_{i L}^l} \R \psi_{iR}^r +y_\psi^{3}  \overline{ \psi^{\rm D}_L }  \R \psi^{\rm D }_R  + y_H^{1} \overline{\psi^{\rm D}_L} H \psi_{2R}^r +y_H^2 \overline{\psi^{\rm D}_L } \tilde{H} \psi_{1R}^r  \nonumber \\ &&+ y_H^3 \overline{\psi_{1L}^l} H^T \psi^{\rm D}_R + y_H^4 \overline{\psi_{2L}^l} H^\dagger \psi_R^{\rm D}  + {\rm h.c.}  \label{yukawa}
\end{eqnarray} 
which give rise to masses of  new fermions as the $U(1)_X$ symmetry is spontaneously broken. 
Yukawa couplings of new fermions with the SM Higgs are constrained by the electroweak precision measurements (EWPM), which was studied in Ref.~\cite{Chao:2016avy}. 
In this paper we assume these Yukawa couplings($y_H^2, ~y_H^3$ and $y_H^4$) are negligible for simplicity.  
 In this case there is no constraint of the EWPM and no extra contribution to effective couplings of $hgg$ and  $h\gamma \gamma$ vertices\footnote{
We refer the reader to Ref.~\cite{Chao:2014dpa} for the study of ${\rm BR} (h\to \gamma \gamma)$ arising from VL fermions, which is similar to our case.  The contribution of VL fermions is negligible for the small Yukawa coupling scenario.},  where $h$ is the SM Higgs.
The Yukawa interaction between the SM quarks and new fermions is forbidden by the $U(1)_X$ symmetry.
Thus there is no mixing between the new colored states and SM quarks. 
%
%
To avoid the problem of the long lived charged/colored particle, one needs to introduce a flavored scalar dark matter $\chi$,  whose $U(1)_X$ charge is $2/3$. $\chi$ couples to  $\psi_{2L}^l$ and  the right-handed top quark:
\begin{eqnarray}
{\cal L}_{\rm DM}^{} \supset  c_\chi \overline{\psi_{2 L}^l} \chi  t_R + {\rm h.c.}.
\end{eqnarray}
The decay chain  of  charged fermions is then $ \psi_{Q=5/3} \to \psi_{Q=2/3}^\prime +W^+ \to \chi + t+ W^+ $.
In this way, constraints from searching new color states  at the LHC can be greatly loosed, which is somehow similar to the case of the stealth supersymmetry~\cite{Fan:2011yu}.
Notice that $\chi$ is automatically stabilized by the $U(1)_X$ symmetry in this case.
We refer the reader to Ref.~\cite{Chao:2016avy} for the detail of the dark matter phenomenology in the gauged B+L symmetry.  

The gauge interaction of $\R$ as well as the scalar potential  can be written as
\begin{eqnarray}
{\cal L} \ni  ( D_\mu \U )^\dagger (D^\mu \U ) - \left\{ -\mu^2 H^\dagger H+ \lambda (H^\dagger H)^2 - \mu_1^2 \U^\dagger \U + {\lambda_1} (\U^\dagger \U)^2  + \tilde\lambda (\U^\dagger \U) (H^\dagger H) \right\}
\end{eqnarray}
where  $D_\mu = \partial_\mu - i Y_X  g_X Z_\mu^{\prime} $ with $Z^\prime_\mu$ the gauge field of the $U(1)_X$, $\U=(\R+v_\R +i A_\R)/\sqrt{2}$. The potential  contains no CP violation but results in the $\R-H$ mixing through the term: $\tilde\lambda (\R^\dagger \R ) (H^\dagger H)$.    
According to the minimization conditions one has
\begin{eqnarray}
v_{}^2= {4 \lambda_1 \mu^2 -2\tilde \lambda \mu_1^2 \over 4 \lambda  \lambda_1 -\tilde\lambda^2} \; , \hspace{2cm} v_\R^2 =  {4 \lambda \mu_1^2 -2\tilde \lambda \mu^2 \over 4 \lambda  \lambda_1 -\tilde\lambda^2} \; ,
\end{eqnarray}
where $v$ and $v_\R$ are the VEVs of the SM Higgs and $\U$ respectively.
As $\U$ gets non-zero VEV the $U(1)_X$ gauge symmetry is spontaneously broken.
The mass eigenvalue of $Z^\prime $ is then $ M_{Z^\prime} \approx 2 b  g_X^{}v_\R$.  
Notice that $\R$ might also interact with the dark matter $\chi$, whose effect will be discussed in section IV.

\section{Constraints}

After the spontaneous breaking of the electroweak and the $U(1)_X$ symmetries, the CP-even scalar mass matrix in the basis $(h, ~\R)$ can be written as
\begin{eqnarray}
M_{\rm CP~even}^2 =\left( \matrix{ 2 \lambda v^2 & \tilde \lambda v v_S \cr \tilde \lambda v v_S & 2 \lambda_1 v_S^2 } \right) \; , \label{massmatrix}
\end{eqnarray}
which can be diagonalized by the $2\times 2 $ unitary transformation.  Physical parameters of this model are then $m_h$, $m_{\R}$, $\theta$, $v_\R $, $y_\psi^{i}$ (i=1,2,3) and $m_{Z^\prime}$, where $\theta$ is the mixing angle of $\R$ with the SM Higgs.  
Mass eigenvalues of new fermions can be roughly written as $m_{\psi}^i  = y_\psi^i v_\R  $, and parameters in the potential can be reconstructed from the squared mass matrix (Eq.~(\ref{massmatrix})) in terms of  mass eigenvalues, mixing angle $\theta$ and VEVs:
\begin{eqnarray}
\tilde \lambda~ &=& (m_\R^2-m_h^2) cs {1\over v v_\R} \label{lambda1} \; , \\
\lambda~ &=& {1\over 2 v^2 } \left[ m_h^2 c^2 + m_\R^2 s^2  \right] \label{lambda2}  \; , \\
\lambda_1 &=& {1\over 2 v_s^2 } \left[ m_h^2 s^2 + m_\R^2 c^2 \right]  \label{lambda3} \; ,
\end{eqnarray}
where $c= \cos \theta$ and $ s=\sin \theta$.  
%
%
$\mu^2$ and $\mu_1^2$ can be determined by tadpole conditions. The decay rate of $\R$  can be written as
\begin{eqnarray}
\Gamma (\R \to 2 \psi ) &= & { c^2 n_C m_\psi^2 (m_\R^2 - 4 m_\psi^2 )^{3/2} \over  8 \pi  m_\R^2 v_\R^2 } \theta(m_R-2 m_\psi) \; , \\
\Gamma (\R \to 2 V ) &= & {s^2 \sqrt{m_\R^2 - 4 m_V^2 } \over (1+\delta_V )4 \pi v^2 m_\R^2} \left(3 m_{V}^4 - m_{\R}^2 m_{V}^2 + {1\over 4 } m_{\R}^4  \right) ~~(V=W,Z) \; , \\
\Gamma (\R \to t\bar t )~ &= & { s^2 n_C m_t^2 (m_\R^2 - 4 m_t^2 )^{3/2} \over  8 \pi  m_\R^2 v^2 } \; , \\
\Gamma (\R \to hh ) &\approx & {\sqrt{m_\R^2 -4 m_h^2 } \over 32 \pi m_\R^2 } \left| \tilde\lambda v_s c^3 +6 cs^2  ( \lambda_1 v_\R - \lambda v )\right|^2 \; .
\end{eqnarray}
where $n_C$ is the color number, $\delta_W=0$ and $\delta_Z=1$\footnote{ The reason we define $\delta_Z=1$ and $\delta_W=0$ is that there are two identical particles in the $ZZ$ channel, such that $\Gamma(\R \to ZZ) $ is phase space suppressed compared with $\Gamma (\R \to WW)$.   }, $m_V$ $(V=W/Z)$ represents the mass of  $W$ or $Z$ gauge boson, $m_\psi$ is the mass of new colored fermion.
In our model $\R$ couples to the SM fermions via its mixing with the SM Higgs. 
Such that its decay rates to the SM fermions equal to these of the SM-like Higgs multiplied by the factor $\sin^2\theta$. 
The decay rate of $\R \to \bar u u$ is then suppressed by both the mixing angle and the tiny Yukawa coupling of the $u$ quark with the SM Higgs. 
As a result, the rate of $\R \to \bar u u $ can be safely neglected. 
For interactions with vector bosons, $\R$ couples to $WW$ and $ZZ$ through the mixing with SM Higgs and couples to the $Z^\prime Z^\prime$ directly due to its nonzero $U(1)_X$ charge, which is the typical feature of our model compared with the case of CP-odd scalar, etc..
Notice that the value of $\sin \theta$ plays very important rule in the decay of $\R$,  however  it  is constrained by the Higgs measurements.  
Performing a universal Higgs fit~\cite{Giardino:2013bma} to the current data given by the ATLAS and CMS collaborations, one gets the  bound on the mixing angle, which has $c>0.865$~\cite{weichaodm} at the 95\% confidence level.
For the constraint of the oblique parameters~\cite{Peskin:1991sw},  restrictions from the $S$ and $T$ parameters are negligible   because of the near  degeneracy of the vector-like fermions, as can be seen from Eq. (\ref{yukawa}).
In our model  $\R$ couples  to photon and gluon pairs at the one-loop level, with new colored fermions running in the loop.  
Decay rates of $\R$ to $gg$ and $\gamma \gamma$ can be written as 
\begin{eqnarray}
\Gamma (\R \to \gamma \gamma ) &\approx &  {\alpha^2 m_\R^3 \over 1024  \pi^3 v_\R^2 }   \left| \sum_\psi  2 n_C Q_\psi^2A_{1/2} (\tau_\psi) \right|^2  \; ,\\
\Gamma (\R \to gg ) &\approx &  {\alpha_s^2 m_\R^3 \over 128  \pi^3 v_\R^2 } \left|\sum_\psi A_{1/2} (\tau_\psi) \right|^2  \; .
\end{eqnarray} 
where $ \tau_\psi = 4 m_\psi^2 /m_\R^2 $.  the expression of the loop function $A_{1/2} (x) $ can be found in~Refs.~\cite{Chao:2014dpa,Chao:2015ttq}.  
%
%
%

\begin{figure}
  \includegraphics[width=0.45\textwidth]{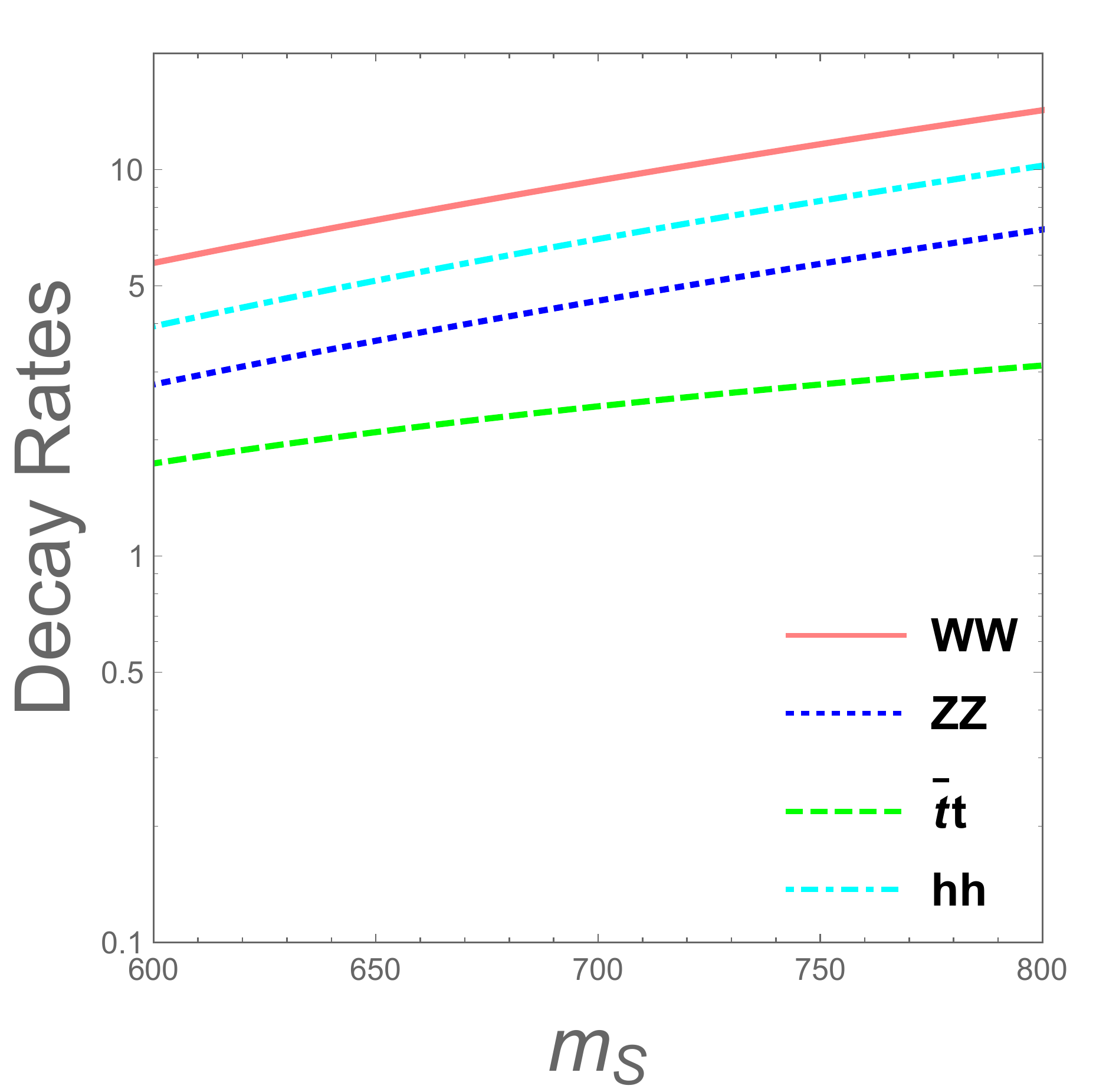}
  \includegraphics[width=0.45\textwidth]{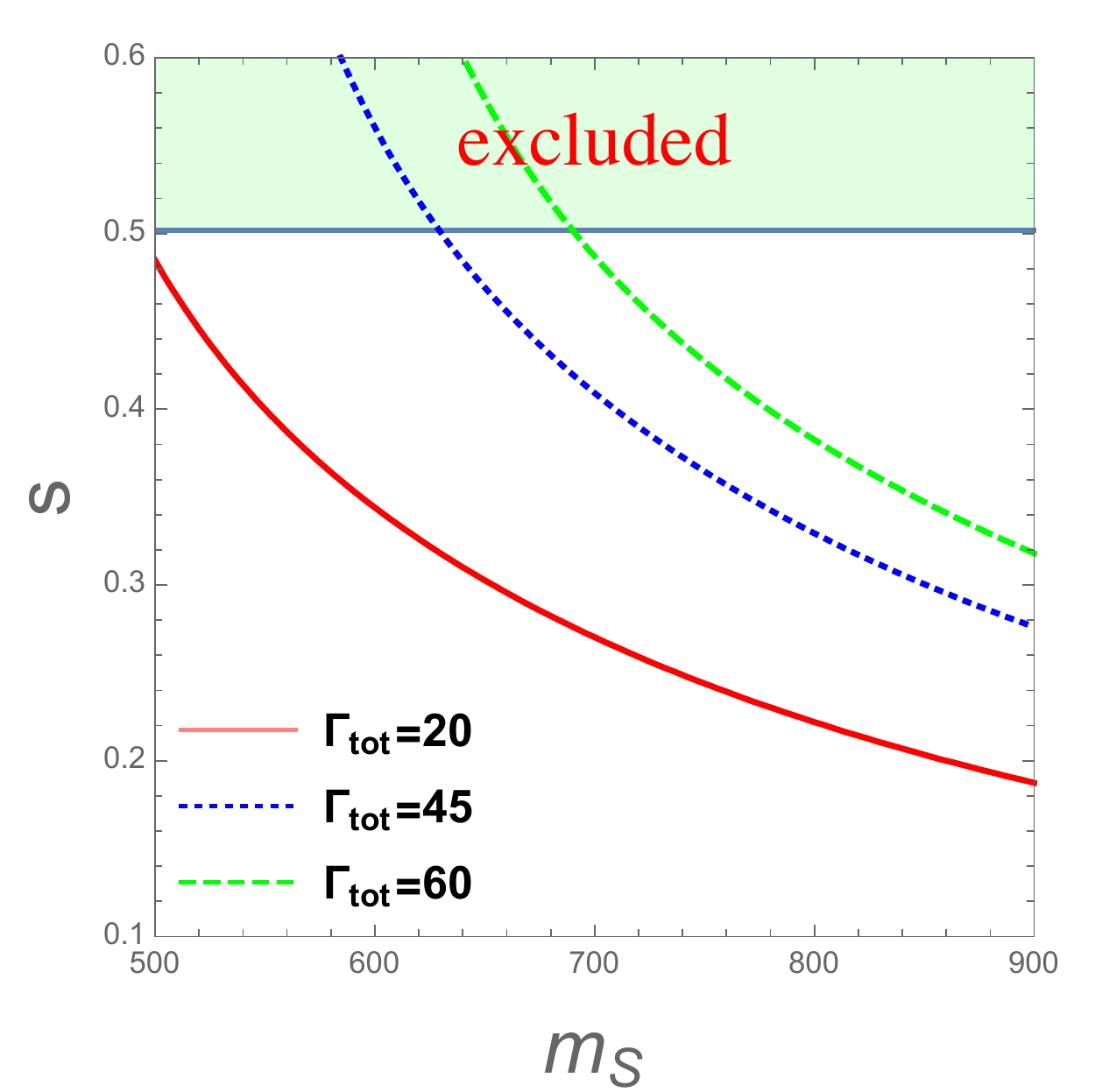}
\caption{\label{rate}   Left panel:  Decay rates of $\R$ as the function of $m_\R$ by setting $s=0.4$; Right panel: Contours of the total decay rate in the $m_\R -s$ plane, where the green region is excluded by the Higgs measurements.}
\end{figure}

\begin{figure}
  \includegraphics[width=0.45\textwidth]{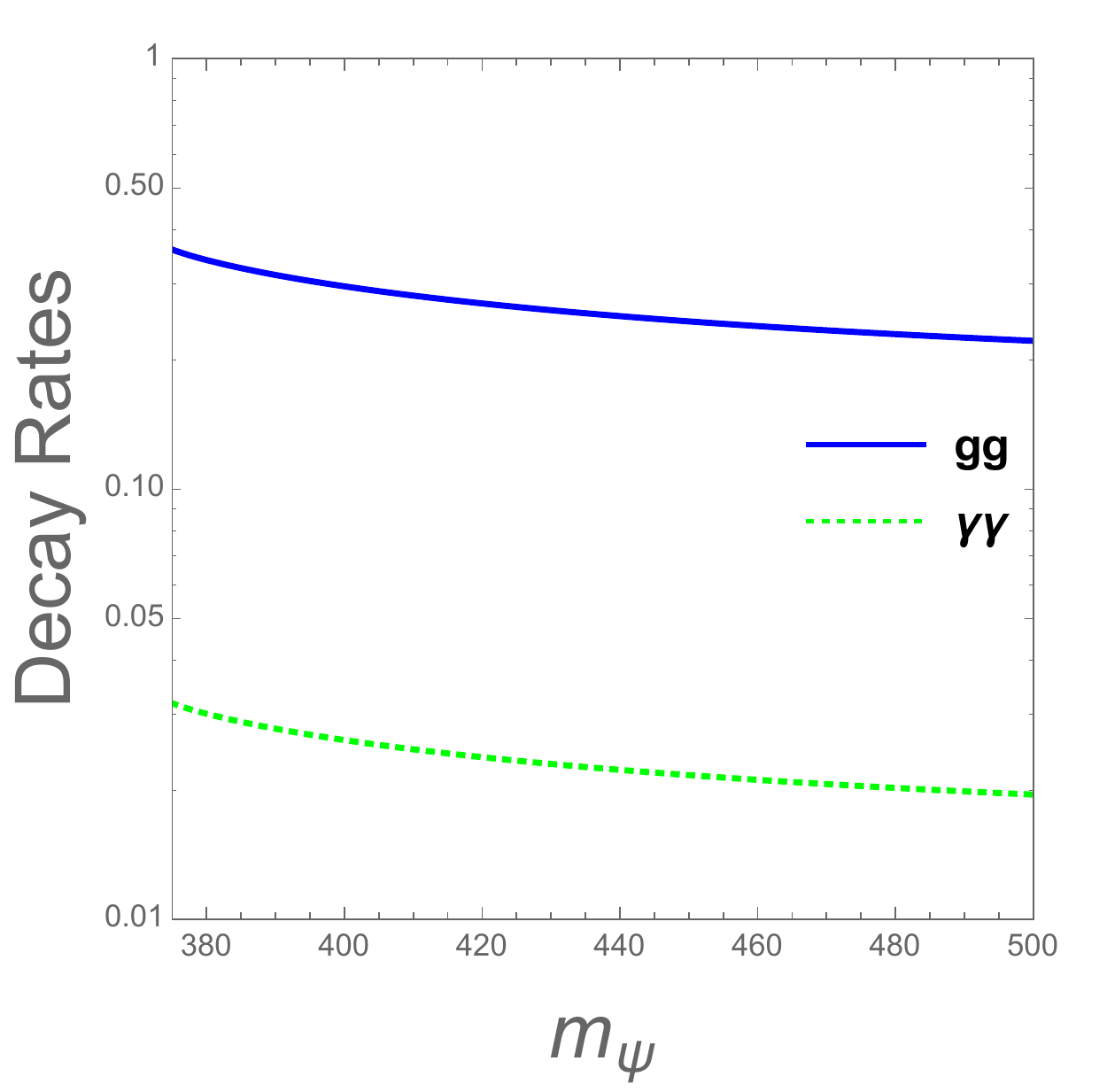}
  \includegraphics[width=0.45\textwidth]{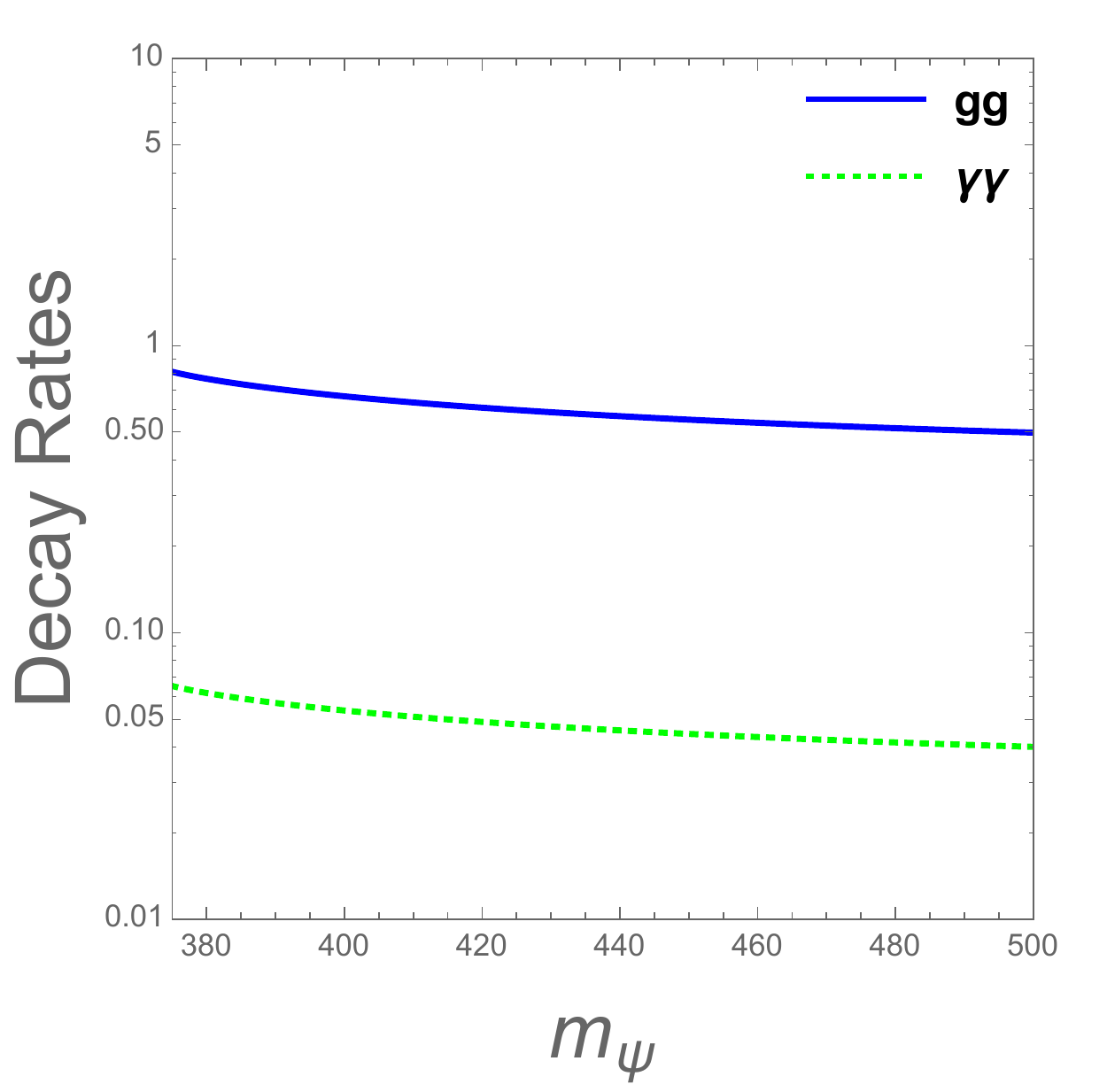}
\caption{\label{rategg}  Decay rate as the function of new fermion masses the $gg$ and $\gamma \gamma$ channels,  by setting $v_\R=400$ GeV and  $m_\R=750~{\rm GeV}$. Left panel corresponds to the minimal particle contents showed in the Table.~\ref{aaa}, while right panel corresponds to the case of adding two extra vector like quarks.  }
\end{figure}

Assuming $ m_\R < 2 m_{Z^\prime ,\psi}$,  $\R$  might mainly decay into the SM final states though its mixing with the SM-like Higgs or through the triangle loop. 
We show in the left panel of Fig.~\ref{rate} the decay rates of $\R$ as the function of $m_\R$ in $WW,~ZZ,\bar t t$ and $hh$ channels, by setting $v_\R=400~{\rm GeV}$ and $s=0.4$.  
It shows that $WW$ and $hh$ channels dominate the decay of $\R$. 
The decay rate of $\R$ in the $hh$ channel will overtake the decay rate of $\R$ in the $WW$ channel as $v_\R<300~{\rm GeV}$.  
We show in the right panel of Fig.~\ref{rate} contours of the total decay rate in the $m_\R-s$ plane, where the solid, dotted and dashed lines correspond to $\Gamma_{\rm tot}=20, ~45,~60$ GeV respectively and the green region is excluded by the Higgs measurements.
Taking $m_\R=750~{\rm GeV}$, one gets $\Gamma_{\rm tot} =45~{\rm GeV}$ for $s=0.365$ which is consistent with all current bounds. 
Since decays of $\R$ into diphoton and gluon-gluon final states can only happen at the one-loop level, their rates are small compared with the tree-level processes, and are sensitive to the electric charges and colors of the particle in the triangle loop.  
We show in the left panel of  Fig. \ref{rategg}  decay rates of $\R$ as the function of vector-like quark mass in the $gg$ and $\gamma \gamma$ channels,  by setting $v_\R=400$ GeV,  $m_\R=750~{\rm GeV}$ and  assuming a degenerate vector-like quark masses for simplicity.  
In this case, the ratio  $ r =\Gamma(\R\to gg )/ \Gamma(\R \to \gamma \gamma ) \sim 11.5 $, which is the typical feature of this model. 
Actually, the diphoton decay rate can be greatly enhanced by choosing a relatively larger  weak hypercharge  or adding new vector like quark pairs which cancel anomalies automatically and might have Yukawa interactions with the $\R$.  
 We plot in the  right panel of Fig. \ref{rategg}, decay rates of $\R$ for the case where there is two more vector-like fermions $\psi_{(3L,3R)}$ and $\psi_{(4L,4R)}^{}$, where quantum numbers of $\psi_{(3L,3R)}^{}$ are the same as $\psi_{(1L,1R)}^{(l,r)}$ while $\psi_{(4L,4R)}$ have opposite $U(1)_X$ charges compared with that of  $\psi_{(3L,3R)}$\footnote{In this scenario the Eq. (\ref{master}) does not change. }, and extra fermions have the same Yukawa coupling with  $\R$.  
Decay rates are greatly enhanced in this case.
$\R$ might also decay into dark matter pair, whose rate is proportional to the coupling of $\R$ to the dark matter, which is  a little bit arbitrary and whose effect will be studied in the next section.  
The phenomenology of the dark matter arising from this model, which is interesting but beyond the reach of this paper, will be shown in a future study~\cite{Chao:2016avy}.

\begin{figure}
  \includegraphics[width=0.45\textwidth]{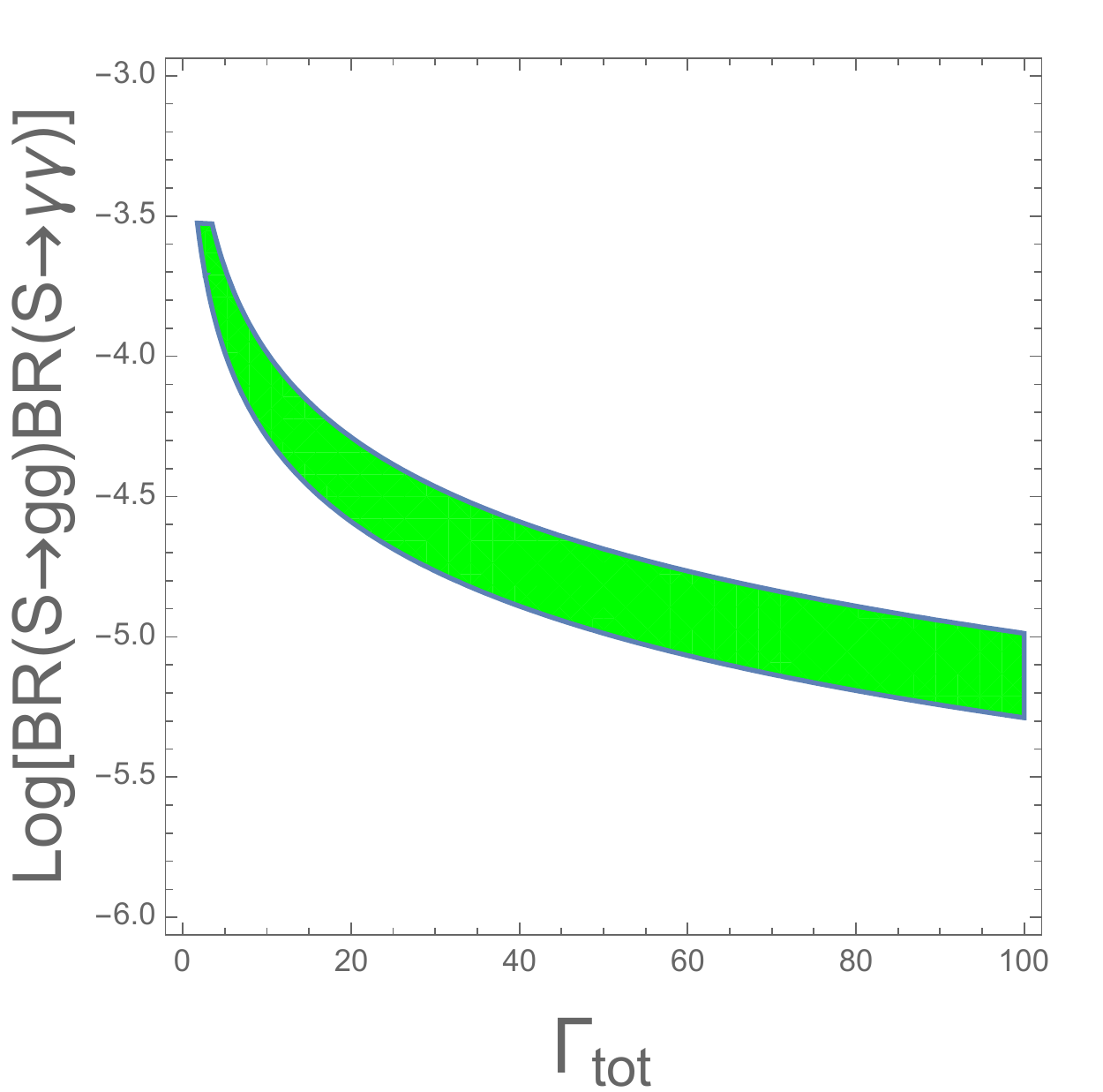}
\caption{\label{totrate}   Region in the $\Gamma_{\rm tot}^{\R} - {\rm Log} [{\rm BR } (\R \to gg ) \times {\rm BR } (\R \to \gamma \gamma )]$ plane, that has $\sigma (pp \to \R \to \gamma \gamma ) \approx 5 \sim 10 ~{\rm fb}$.  }
\end{figure}

\section{Diphoton excess}

Both the CMS and ATLAS collaborations have observed the diphoton excess at $m_{\gamma \gamma} =750~{\rm GeV}$, whose cross section can  be roughly estimated as  $\sigma (pp \to \R \to \gamma \gamma ) \approx 5 \sim 10 ~{\rm fb}$. 
The best fit width of the resonance is about 45 GeV. It is interesting to interpret this result as a signal of new physics.  
But it is not trivial to find the new physics that can reasonably fit with the observed data.  
The production cross section of the resonance at the LHC is given in Eq. (\ref{basic}), with the numerical value of partonic integral evaluated at $M=750~{\rm GeV}$ and $\sqrt{s}=13~{\rm TeV}$ as $C_{gg} \approx 3163$~\cite{Franceschini:2015kwy}. 
We plot in Fig. \ref{totrate} the region in the $\Gamma_{\rm tot}^{\R} - {\rm Log} [{\rm BR } (\R \to gg ) \times {\rm BR } (\R \to \gamma \gamma )]$ plane that might give rise to the observed cross section.  
For our model, we take the following two group benchmark inputs of $\Gamma (\R \to gg )$ and $\Gamma (\R \to \gamma \gamma )$, based on the numerical simulations shown in the left and right panels of Fig.  \ref{rategg} by setting $v_\R =m_\psi =400~{\rm GeV}$\footnote{ The collider signature of $\psi$ in our model is $pp\to \psi \bar \psi \to jj+\chi\chi$(missing energy). For the case $\psi$ being only slightly heavier than $\chi$, the dijet will be very soft and $m_\psi=400~\text{GeV}$ will be compatible with the LHC direct searches of VL quarks, which is similar to the case of the stealth supersymmetry~\cite{Fan:2011yu} and stealth top model~\cite{Chao:2015ttq}. },
\begin{eqnarray}
&& (\mathbf{Benchmark~ I}~) \hspace{0.5cm } \Gamma(\R\to gg ) = 0.296~{\rm GeV} \; , \hspace{0.3cm} \Gamma(\R \to \gamma \gamma ) = 0.026~{\rm GeV}\; ; \\
&& (\mathbf{Benchmark~ II}) \hspace{0.5 cm } \Gamma(\R\to gg ) =  0.665~{\rm GeV}\; , \hspace{0.3cm} \Gamma(\R \to \gamma \gamma ) = 0.054~{\rm GeV}\; .
\end{eqnarray}
We derive the benchmark I using the particle contents given in Table. \ref{aaa}, while the benchmark II corresponds to the case  of  extending the particle contents of the benchmark I with two extra vector-like fermions $\psi_{(3L,3R)}$ and $\psi_{(4L,4R)}$ without introducing any further anomalies.

\begin{figure}
  \includegraphics[width=0.45\textwidth]{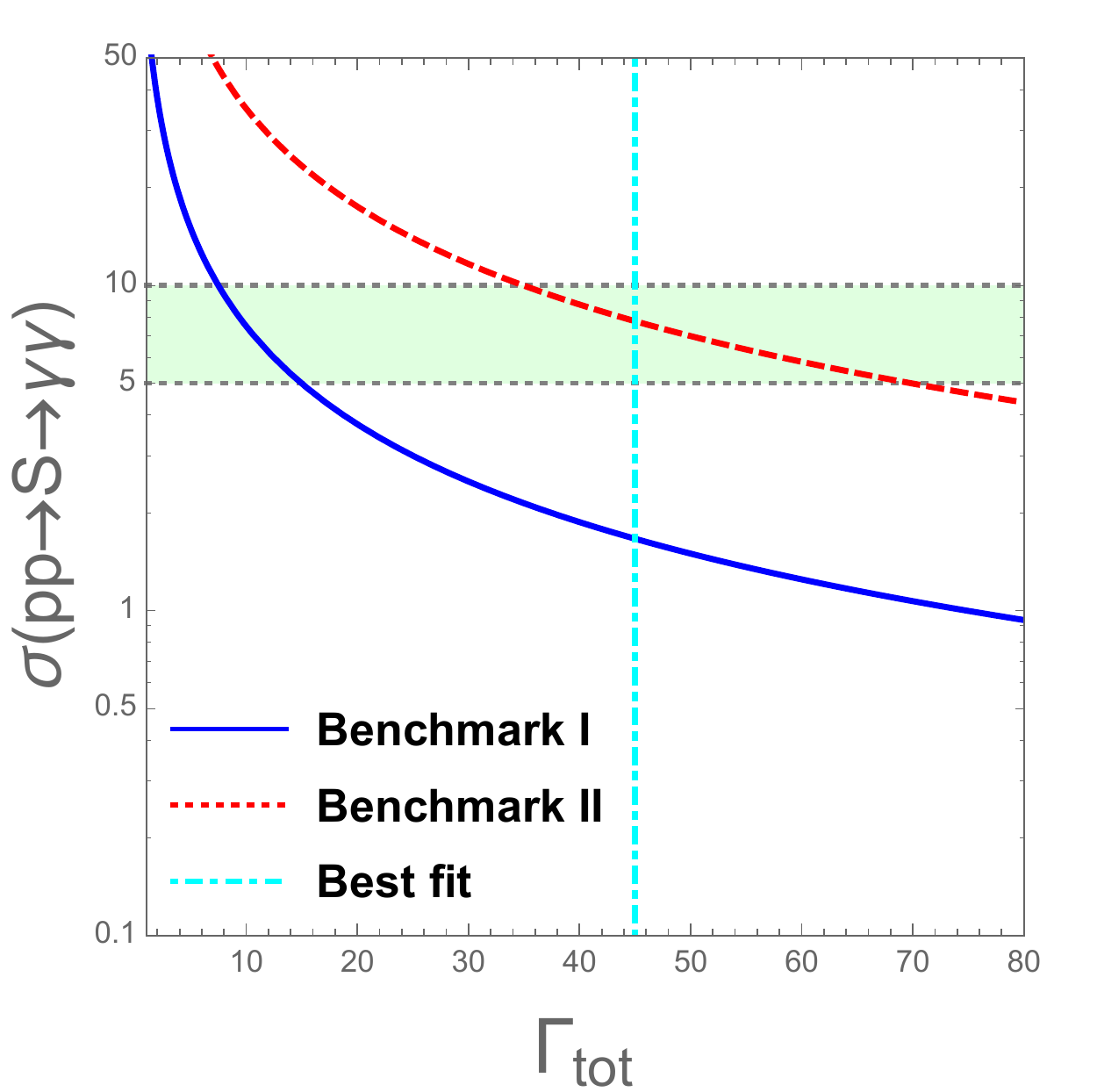}
\caption{\label{benchmark}   Production cross section as the function of $\Gamma_{\rm tot}^\R$ for the benchmark (I) and (II).  }
\end{figure}

Using these two benchmark inputs we plot the production cross section of $\R$ as the function of  $\Gamma_{\rm tot}^{\R} $  in Fig.~\ref{benchmark}.   
The solid and dotted lines correspond to the prediction of benchmark I and benchmark II respectively, the vertical dot-dashed is the best fit value of the total decay rate, while the light green band between the grey horizontal lines is the observed cross section.
It shows that $\Gamma_{\rm tot}^\R \in (7.5,~15)$ for benchmark I and $\Gamma_{\rm tot}^\R \in (35, 70)$ for benchmark II. 
Taking  $\Gamma_{\rm tot}^\R$ as its best fit value ($45$ GeV from the ATLAS), one has $\sigma (pp\to \R \to \gamma \gamma ) = 1.7, ~7.8 $ fb for benchmark I and benchmark II respectively.  
Benchmark II provides better explanation to the data, but benchmark I is still available up to upon a more precise measurement of the total rate.    
As was discussed in the last section, $\R$ might decay into  SM final states through its mixing with the SM Higgs boson, or dark matter pair through the interaction $\lambda_{\rm DM} (\U^\dagger \U) \chi^\dagger \chi$.  The $\Gamma(\R\to {\rm invisible})$ can be written as  
\begin{eqnarray}
\Gamma(\R \to {\rm invisible} ) \approx { (\lambda_{\rm DM} v_\R)^2 \over 16 \pi m_\R^2 }  \sqrt{m_\R^2 -4 m_\chi^2 } \; .
\end{eqnarray}
This rate depends on a new parameter $\lambda_{\rm DM}$, which is somehow arbitrary in this model. 
If this rate is negligible, one might get the bound on the mixing angle $\theta$ from the observed cross section:  $\sin \theta \in[0.15,~0.21]$ for benchmark I and $\sin \theta\in [ 0.32,~0.50]$ for benchmark II. 
These constraint are still consistent with the constraint of Higgs measurements. 
On the other hand, the Run-1 LHC at the 8 TeV has searched for the diHiggs, $WW$, $ZZ$ and $\bar t t$ final states, and  the non-observation any signal puts upper bound  on these cross sections: ${\rm \sigma\cdot BR(ZZ) <12}$ fb~\cite{Aad:2015kna},     ${\rm \sigma\cdot BR(WW) <40}$ fb~\cite{Aad:2015agg}, ${\rm \sigma\cdot BR(hh) <39}$ fb~\cite{Aad:2015xja} and ${\rm \sigma\cdot BR(\bar t t) <550}$ fb~\cite{ttbarx}.
Each channel puts a upper bound on the $\sin\theta$, it turns out the strongest constraint comes from the $ZZ$ channel, which has $\sin\theta< 6\times 10^{-3}$.  
In this case $\Gamma(\R \to {\rm SM~final~states})$ is negligible and  the $\R\to \chi \chi$ channel dominates the contribution to the total width. One has $\lambda_{\rm DM } \in (1.3,~1.9)$  for the benchmark I and $\lambda_{\rm DM } \in (2.8,~4.0)$ for the benchmark II. 
It should be mentioned that $\R$ may decay into the new fermion final states when $m_\R>2m_\psi$. 

Finally let us consider the constraint of the $Z^\prime$.
A heavy $Z^\prime $, whose couplings with SM fermions are the same as these of $Z$ boson,  was searched at the LHC in the dilepton channel, which is excluded at the 95\% CL for $M_{Z^\prime } <2.9~{\rm TeV}$~\cite{Aad:2014cka} and for $M_{Z^\prime } <2.79~{\rm TeV}$~\cite{Khachatryan:2014fba}.
To be consistent with this constriant, one extra scalar singlet with the same quantum number as $\R$ should be introduced, whose VEV, $v_{\rm new}$, breaks the new gauge symmetry spontaneously and contributes to the mass of $Z^\prime$.  
It is similar to the case of two Higgs doublet model of Type-I, where the second Higgs doublet contributes to the electroweak symmetry breaking but has no effect in generating fermion masses. 
In our case $v_{\rm new} $ should be roughly larger than $5.6~{\rm TeV}$ to satisfy the current collider constraint.
The scenario of breaking a local $U(1)_{B/(B+L)}$ symmetry with multi-singlets, which is brand new but beyond the reach of this paper, will be investigated in another project.

\section{Conclusion}

Both the ATLAS and CMS collaborations have observed excesses in the $(pp\to \gamma \gamma)$ channel at the $\sqrt{s} =13 ~{\rm TeV}$.  
If confirmed, it would be very interesting and  important to investigate the new physics behind this excess since no SM process could generate this excess. 
We provide a complete theoretical explanation to  the excess, which is a CP-even scalar, $\R$, that triggers the spontaneous breaking of a local $U(1)_B$ or $U(1)_{B+L}$ symmetry.   
$\R$ couples to gluon and photon pairs at the one-loop level with vector like quarks running in the loop, which are introduced to cancel anomalies of this new $U(1)_X$ gauge symmetry.  
The model also include a scalar dark matter candidate, stabilized by the $U(1)_{B/B+L}$ symmetry,  which couples to new fermions so as to avoid the problem of long lived charged/colored particles.
Our study shows that  this model can naturally explain the diphoton excess without conflicting with other observables.  Notably it (benchmark II) also favor the best fit width given by the ATLAS.  
Constraints on couplings of $\R$ were studied, which showed that $\R$ has a negligible mixing with the standard model Higgs boson, but sizable coupling with the dark matter. 
We expect the run-2 LHC at  high luminosity could shed light on the total width of this new resonance. 
We just simply assume a very much heavy $Z^\prime$, the phenomenology of which,  although important but beyond the reach of this paper, will be given in a future study.   

\begin{acknowledgments}
The author thanks to Huai-ke Guo, Ran Huo, Hao-lin Li, Grigory Ovanesyan  and especially Jiang-hao Yu for very helpful discussions. 
This work was supported in part by DOE Grant DE-SC0011095.
\end{acknowledgments}


\begin{thebibliography}{99}


\bibitem{ATLAS13}
The ATLAS Collaboration, ``Search for resonances decaying to photon pairs in 3.2 fb$^{-1}$ of pp collisions at $\sqrt{s}=13$ TeV with the ATLAS detector'', Tech. Rep. ATLAS-CONF-2015-081, CERN, Geneva, Dec, 2015.

\bibitem{CMS13}
  The CMS Collaboration, ``Search for new physics in high mass diphoton events in proton-proton collisions at 13 TeV'', Tech. Rep. CMS-PAS-EXO-15-004, CERN, Geneva, 2015.


\bibitem{Landau:1948kw} 
  L.~D.~Landau,
  Dokl.\ Akad.\ Nauk Ser.\ Fiz.\  {\bf 60}, no. 2, 207 (1948).
  doi:10.1016/B978-0-08-010586-4.50070-5
  
  
  
\bibitem{Yang:1950rg} 
  C.~N.~Yang,
  Phys.\ Rev.\  {\bf 77}, 242 (1950).
  doi:10.1103/PhysRev.77.242
  


\bibitem{ATLAS8}
  G.~Aad {\it et al.} [ATLAS Collaboration],
  Phys.\ Rev.\ D {\bf 92}, no. 3, 032004 (2015)
  doi:10.1103/PhysRevD.92.032004
  [arXiv:1504.05511 [hep-ex]].

\bibitem{CMS8}
  V.~Khachatryan {\it et al.} [CMS Collaboration],
  Phys.\ Lett.\ B {\bf 750}, 494 (2015)
  doi:10.1016/j.physletb.2015.09.062
  [arXiv:1506.02301 [hep-ex]].
  
  
\bibitem{Franceschini:2015kwy}
  R.~Franceschini {\it et al.},
  arXiv:1512.04933 [hep-ph].

\bibitem{Harigaya:2015ezk}
  K.~Harigaya and Y.~Nomura,
  arXiv:1512.04850 [hep-ph].

\bibitem{Mambrini:2015wyu}
  Y.~Mambrini, G.~Arcadi and A.~Djouadi,
  arXiv:1512.04913 [hep-ph].

\bibitem{Backovic:2015fnp}
  M.~Backovic, A.~Mariotti and D.~Redigolo,
  arXiv:1512.04917 [hep-ph].

\bibitem{Angelescu:2015uiz}
  A.~Angelescu, A.~Djouadi and G.~Moreau,
  arXiv:1512.04921 [hep-ph].

\bibitem{Nakai:2015ptz}
  Y.~Nakai, R.~Sato and K.~Tobioka,
  arXiv:1512.04924 [hep-ph].

\bibitem{Knapen:2015dap}
  S.~Knapen, T.~Melia, M.~Papucci and K.~Zurek,
  arXiv:1512.04928 [hep-ph].

\bibitem{Buttazzo:2015txu}
  D.~Buttazzo, A.~Greljo and D.~Marzocca,
  arXiv:1512.04929 [hep-ph].

\bibitem{Pilaftsis:2015ycr}
  A.~Pilaftsis,
  arXiv:1512.04931 [hep-ph].



\bibitem{DiChiara:2015vdm}
  S.~Di Chiara, L.~Marzola and M.~Raidal,
  arXiv:1512.04939 [hep-ph].

\bibitem{Higaki:2015jag}
  T.~Higaki, K.~S.~Jeong, N.~Kitajima and F.~Takahashi,
  arXiv:1512.05295 [hep-ph].

\bibitem{McDermott:2015sck}
  S.~D.~McDermott, P.~Meade and H.~Ramani,
  arXiv:1512.05326 [hep-ph].

\bibitem{Ellis:2015oso}
  J.~Ellis, S.~A.~R.~Ellis, J.~Quevillon, V.~Sanz and T.~You,
  arXiv:1512.05327 [hep-ph].

\bibitem{Low:2015qep}
  M.~Low, A.~Tesi and L.~T.~Wang,
  arXiv:1512.05328 [hep-ph].

\bibitem{Bellazzini:2015nxw}
  B.~Bellazzini, R.~Franceschini, F.~Sala and J.~Serra,
  arXiv:1512.05330 [hep-ph].

\bibitem{Gupta:2015zzs}
  R.~S.~Gupta, S.~J?ger, Y.~Kats, G.~Perez and E.~Stamou,
  arXiv:1512.05332 [hep-ph].

\bibitem{Petersson:2015mkr}
  C.~Petersson and R.~Torre,
  arXiv:1512.05333 [hep-ph].

\bibitem{Molinaro:2015cwg}
  E.~Molinaro, F.~Sannino and N.~Vignaroli,
  arXiv:1512.05334 [hep-ph].
  
  
\bibitem{Dutta:2015wqh} 
  B.~Dutta, Y.~Gao, T.~Ghosh, I.~Gogoladze and T.~Li,
  arXiv:1512.05439 [hep-ph].
  
  
  
\bibitem{Cao:2015pto} 
  Q.~H.~Cao, Y.~Liu, K.~P.~Xie, B.~Yan and D.~M.~Zhang,
  arXiv:1512.05542 [hep-ph].
  
\bibitem{Matsuzaki:2015che} 
  S.~Matsuzaki and K.~Yamawaki,
  arXiv:1512.05564 [hep-ph].
  
\bibitem{Kobakhidze:2015ldh} 
  A.~Kobakhidze, F.~Wang, L.~Wu, J.~M.~Yang and M.~Zhang,
  arXiv:1512.05585 [hep-ph].
  
\bibitem{Martinez:2015kmn} 
  R.~Martinez, F.~Ochoa and C.~F.~Sierra,
  arXiv:1512.05617 [hep-ph].
  
\bibitem{Cox:2015ckc} 
  P.~Cox, A.~D.~Medina, T.~S.~Ray and A.~Spray,
  arXiv:1512.05618 [hep-ph].
  
\bibitem{Becirevic:2015fmu} 
  D.~Becirevic, E.~Bertuzzo, O.~Sumensari and R.~Z.~Funchal,
  arXiv:1512.05623 [hep-ph].
  
\bibitem{No:2015bsn} 
  J.~M.~No, V.~Sanz and J.~Setford,
  arXiv:1512.05700 [hep-ph].
  
\bibitem{Demidov:2015zqn} 
  S.~V.~Demidov and D.~S.~Gorbunov,
  arXiv:1512.05723 [hep-ph].
  
\bibitem{Chao:2015ttq} 
  W.~Chao, R.~Huo and J.~H.~Yu,
  arXiv:1512.05738 [hep-ph].
  
  
\bibitem{Fichet:2015vvy} 
  S.~Fichet, G.~von Gersdorff and C.~Royon,
  arXiv:1512.05751 [hep-ph].
  
  
\bibitem{Curtin:2015jcv} 
  D.~Curtin and C.~B.~Verhaaren,
  arXiv:1512.05753 [hep-ph].
  
  
\bibitem{Bian:2015kjt} 
  L.~Bian, N.~Chen, D.~Liu and J.~Shu,
  arXiv:1512.05759 [hep-ph].
  
  
\bibitem{Chakrabortty:2015hff} 
  J.~Chakrabortty, A.~Choudhury, P.~Ghosh, S.~Mondal and T.~Srivastava,
  arXiv:1512.05767 [hep-ph].
  
  
\bibitem{Agrawal:2015dbf} 
  P.~Agrawal, J.~Fan, B.~Heidenreich, M.~Reece and M.~Strassler,
  arXiv:1512.05775 [hep-ph].
  
  
\bibitem{Chao:2015nac} 
  W.~Chao,
  arXiv:1512.08484 [hep-ph].
  
  
\bibitem{Chao:2016mtn} 
  W.~Chao,
  arXiv:1601.00633 [hep-ph].
  
\bibitem{Chao:2016aer} 
  W.~Chao,
  arXiv:1601.04678 [hep-ph].
  
  
\bibitem{Csaki:2015vek} 
  C.~Csaki, J.~Hubisz and J.~Terning,
  arXiv:1512.05776 [hep-ph].
  
  
  
\bibitem{Falkowski:2015swt} 
  A.~Falkowski, O.~Slone and T.~Volansky,
  arXiv:1512.05777 [hep-ph].
  
  
\bibitem{Aloni:2015mxa} 
  D.~Aloni, K.~Blum, A.~Dery, A.~Efrati and Y.~Nir,
  arXiv:1512.05778 [hep-ph].
  
\bibitem{Bai:2015nbs} 
  Y.~Bai, J.~Berger and R.~Lu,
  arXiv:1512.05779 [hep-ph];
  A.~Alves, A.~G.~Dias and K.~Sinha,
  arXiv:1512.06091 [hep-ph].
  
  
 
\bibitem{Jaeckel:2012yz} 
  J.~Jaeckel, M.~Jankowiak and M.~Spannowsky,
  Phys.\ Dark Univ.\  {\bf 2}, 111 (2013)
  doi:10.1016/j.dark.2013.06.001
  [arXiv:1212.3620 [hep-ph]].
  
\bibitem{Carmi:2012in} 
  D.~Carmi, A.~Falkowski, E.~Kuflik, T.~Volansky and J.~Zupan,
  JHEP {\bf 1210}, 196 (2012)
  doi:10.1007/JHEP10(2012)196
  [arXiv:1207.1718 [hep-ph]].
  
  
  
  
  
  
  
  
\bibitem{Sakharov:1967dj} 
  A.~D.~Sakharov,
  Pisma Zh.\ Eksp.\ Teor.\ Fiz.\  {\bf 5}, 32 (1967)
  [JETP Lett.\  {\bf 5}, 24 (1967)]
  [Sov.\ Phys.\ Usp.\  {\bf 34}, 392 (1991)]
  [Usp.\ Fiz.\ Nauk {\bf 161}, 61 (1991)].
  doi:10.1070/PU1991v034n05ABEH002497
  
  
  \bibitem{globalsu2}

E. Witten, Phys. Lett. B {\bf 177}, 324(1982)

\bibitem{avector1}

S. L. Adler, Phys. Rev. {\bf 177}, 2426(1969).

\bibitem{avector2}

J. S. Bell and R. Jackiw, Nuovo Cimento A {\bf 60}, 47(1969).

\bibitem{avector3}

W. A. Barden, Phys. Rev. {\bf 184}, 1848(1969).

\bibitem{anog1}

R. Delbourgo and A. Salam, Phys. Lett. B {\bf 40}, 381(1972).

\bibitem{anog2}

T. Eguchi and P. G. O. Freund, Phys. Rev. Lett {\bf 37}, 1251(1976).

\bibitem{anog3}

L. Alvarez-Gaume and E. Witten, Nucl. Phys. B {\bf 234}, 269(1984).


\bibitem{Chao:2010mp} 
  W.~Chao,
  Phys.\ Lett.\ B {\bf 695}, 157 (2011)
  doi:10.1016/j.physletb.2010.10.056
  [arXiv:1005.1024 [hep-ph]].

  
\bibitem{FileviezPerez:2010gw} 
  P.~Fileviez Perez and M.~B.~Wise,
  Phys.\ Rev.\ D {\bf 82}, 011901 (2010)
  [Phys.\ Rev.\ D {\bf 82}, 079901 (2010)]
  doi:10.1103/PhysRevD.82.079901, 10.1103/PhysRevD.82.011901
  [arXiv:1002.1754 [hep-ph]].
  
\bibitem{FileviezPerez:2011pt} 
  P.~Fileviez Perez and M.~B.~Wise,
  JHEP {\bf 1108}, 068 (2011)
  doi:10.1007/JHEP08(2011)068
  [arXiv:1106.0343 [hep-ph]].

\bibitem{Duerr:2013dza} 
  M.~Duerr, P.~Fileviez Perez and M.~B.~Wise,
  Phys.\ Rev.\ Lett.\  {\bf 110}, 231801 (2013)
  doi:10.1103/PhysRevLett.110.231801
  [arXiv:1304.0576 [hep-ph]].
  
  
\bibitem{Giardino:2013bma} 
  P.~P.~Giardino, K.~Kannike, I.~Masina, M.~Raidal and A.~Strumia,
  JHEP {\bf 1405}, 046 (2014)
  doi:10.1007/JHEP05(2014)046
  [arXiv:1303.3570 [hep-ph]].
  
  \bibitem{weichaodm}
  W.~Chao, {\it  Hiding the scalar dark matter direct detection}, to appear.
  
\bibitem{Peskin:1991sw} 
  M.~E.~Peskin and T.~Takeuchi,
  Phys.\ Rev.\ D {\bf 46}, 381 (1992).
  doi:10.1103/PhysRevD.46.381
  
\bibitem{Chao:2014dpa} 
  W.~Chao and M.~J.~Ramsey-Musolf,
  JHEP {\bf 1410}, 180 (2014)
  doi:10.1007/JHEP10(2014)180
  [arXiv:1406.0517 [hep-ph]].






\bibitem{Aad:2015kna} 
  G.~Aad {\it et al.} [ATLAS Collaboration],
  arXiv:1507.05930 [hep-ex].

\bibitem{Aad:2015agg} 
  G.~Aad {\it et al.} [ATLAS Collaboration],
  arXiv:1509.00389 [hep-ex].

\bibitem{Aad:2015xja} 
  G.~Aad {\it et al.} [ATLAS Collaboration],
  Phys.\ Rev.\ D {\bf 92}, 092004 (2015)
  doi:10.1103/PhysRevD.92.092004
  [arXiv:1509.04670 [hep-ex]].

\bibitem{ttbarx}
V. Khachatryan {\it et al.}  [CMS Collaboration], CMS-PAS-B2G-12-006.


\bibitem{Aad:2014cka} 
  G.~Aad {\it et al.} [ATLAS Collaboration],
  Phys.\ Rev.\ D {\bf 90}, no. 5, 052005 (2014)
  doi:10.1103/PhysRevD.90.052005
  [arXiv:1405.4123 [hep-ex]].

\bibitem{Khachatryan:2014fba} 
  V.~Khachatryan {\it et al.} [CMS Collaboration],
  JHEP {\bf 1504}, 025 (2015)
  doi:10.1007/JHEP04(2015)025
  [arXiv:1412.6302 [hep-ex]].

\bibitem{Fan:2011yu} 
  J.~Fan, M.~Reece and J.~T.~Ruderman,
  JHEP {\bf 1111}, 012 (2011)
  doi:10.1007/JHEP11(2011)012
  [arXiv:1105.5135 [hep-ph]].
  
\bibitem{Chao:2016avy} 
  W.~Chao, H.~k.~Guo and Y.~Zhang,
  arXiv:1604.01771 [hep-ph].

\end{thebibliography}
\end{document}